\documentstyle[twoside, epsf]{article}

\input ibvs2.sty

\begin{document}
\IBVShead{5xxx}{00 Month 2015}

\IBVStitletl{Call for Follow-Up Observations of the}{Dynamically Changing Triple Star KIC 2835289}

\IBVSauth{CONROY, K.$^{1,2}$; PR\v SA, A.$^2$, STASSUN, K.$^1$; OROSZ, J.$^3$}

\IBVSinst{Vanderbilt University, Department of Physics and Astronomy, Nashville, TN 37235, USA}
\IBVSinst{Villanova University, Dept.~of Astrophysics and Planetary Science, Villanova PA 19085, USA}
\IBVSinst{San Diego State University, Department of Astronomy, San Diego, CA 92182, USA}

\IBVSkey{photometry}
\SIMBADobj{KIC 2835289}

\begintext

KIC 2835289 (2MASS J19072623+3801389; $\alpha_{2000} = 19^h 07^m 26.231^s$, $\delta_{2000} = +38^\circ 01'$ $38.92''$) is a V$\sim$13 ellipsoidal variable with an orbital period of $\sim$0.86 days, observed by the {\sl Kepler} mission (Borucki et al.~2010) and cataloged in the {\sl Kepler} Eclipsing Binary Catalog (Pr\v sa et al.~2011, {\tt http://keplerEBs.villanova.edu}). {\sl Kepler} observations are available in long cadence (30-min exposure times) for all 17 quarters spanning 4 years, and in short cadence (1-min exposure times) for Quarter 17. It is incorrectly classified in the {\sl Simbad} database as an overcontact binary of the W UMa type.

\IBVSfig{6cm}{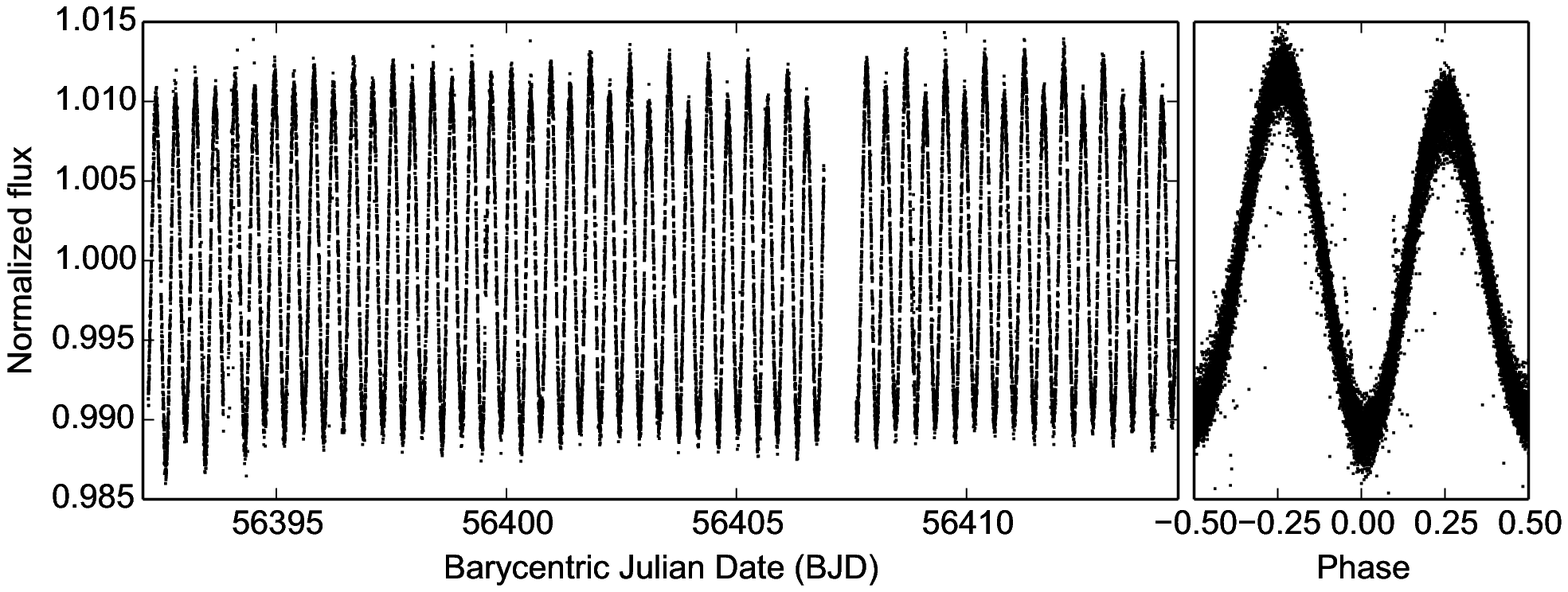}{
{\sl Kepler} light curve of KIC 2835289 in time (left) and phase (right). Chromospheric activity complicates the modeling of this ellipsoidal variable, but with 4 years of uninterrupted data, the intrinsic ellipsoidal signal can be extracted to a high precision.}

Two features make this binary star interesting: strong chromospheric activity and the presence of a third stellar component. Fig.~1 depicts a light curve segment from Quarter 17 in time (left) and phase (right), where the ''smear`` in the phased light curve is a consequence of chromospheric activity. Spots account for the time-varying O'Connell effect -- asymmetric light curve maxima -- that are clearly evident in the data, and interesting in their own right. The remarkable feature of this system, however, is the detection of two eclipse events in quarters Q9 and Q17 (cf.~Fig.~2) separated by $\sim$750 days, attesting to the presence of the third stellar component in this system. We also measured the timings of ellipsoidal minima (akin to eclipse timing variations in eclipsing binary systems; Conroy et al.~2014) and found clear evidence of a moderately eccentric outer orbit (Fig.~3). This circumstance is particularly appealing for further study, since it directly tests the Kozai-Lidov model of close binary evolution. According to the model, three-body systems with eccentric outer orbits undergo oscillations in inner eccentricity and mutual inclination angle (Fabrycky \& Tremaine 2007). While in a state of higher inner eccentricity, tidal friction at periastron increases and circularization ensues (Kiseleva et al.~1998). In consequence, the inner binary is tightened while the outer star settles into a wide, eccentric, inclined orbit, much like KIC 2835289.

\IBVSfig{8cm}{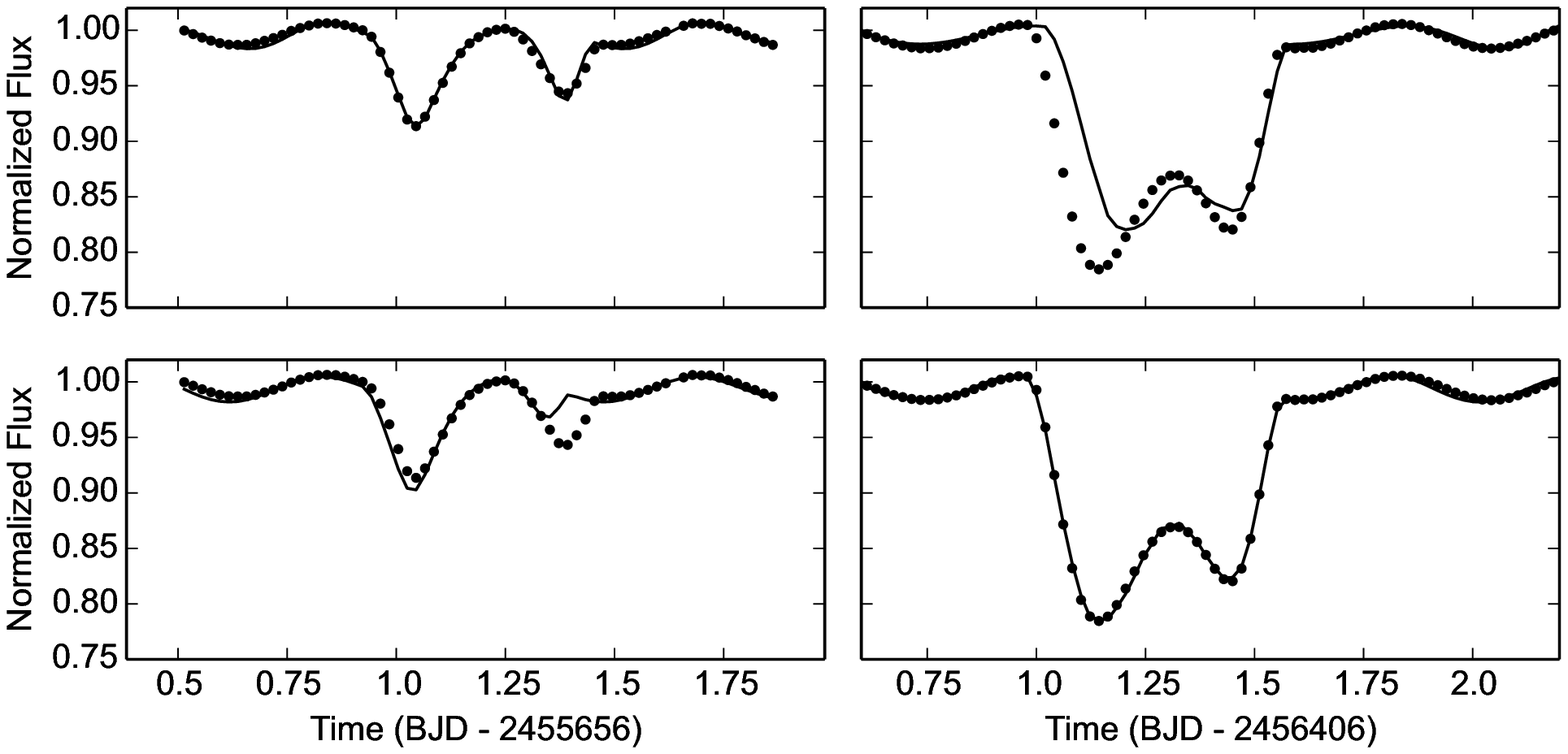}{
Eclipse events in {\sl Kepler} Quarters 9 (left) and 17 (right). Filled circles are long cadence observations and solid lines are model fits. No single model was found to fit both eclipse events, but allowing the inner eccentricity and mutual inclination to vary, a family of solutions was found that can describe observations. The model fit to Q9 data is plotted in the top panels, and the model fit to Q17 data is plotted in the bottom panels.
}

\IBVSfig{8cm}{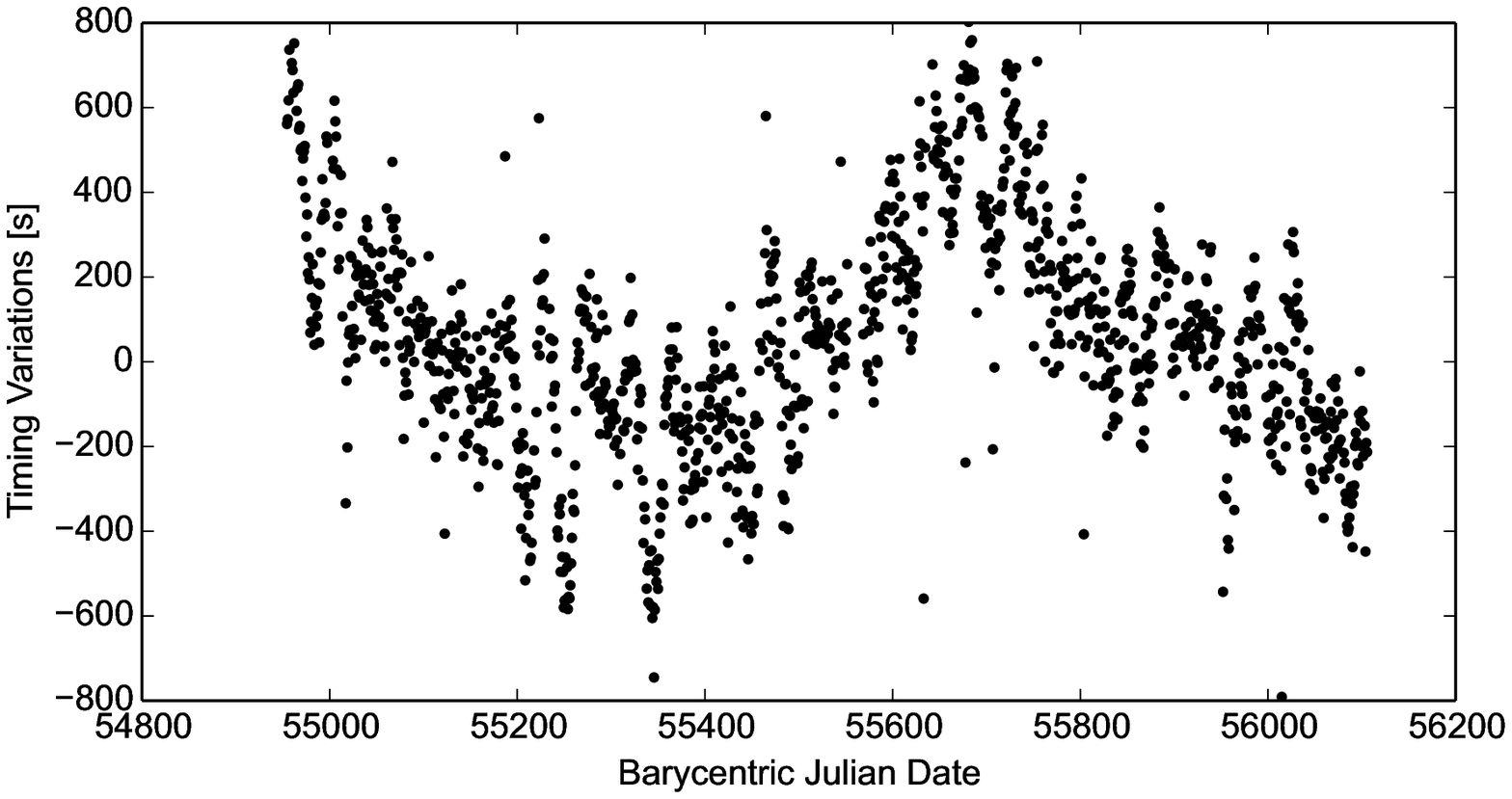}{
Timings of ellipsoidal minima. The dominant trend is the light time travel effect due to third body interaction. Chromospheric activity causes secular variations.}

An ongoing detailed photodynamical analysis (Conroy et al.~2015, in preparation) using a new version of the PHOEBE code (Pr\v sa \& Zwitter 2005, Conroy et al.~2013) provided a uniquely interesting first look into the dynamics in this system. While quite a few triple systems have been successfully modeled recently based on the dynamical effects in timing variations (cf.~Borkovits et al.~2015, who analyzed mutual inclinations of 26 triple stars in the {\sl Kepler} data-set), the near-circular nature of the tight inner binary in this system excluded it from their analysis. We ran the photodynamical model in PHOEBE and found that there seems to be no consistent set of orbital and physical parameters that would successfully describe \emph{both} eclipse events. However, by allowing the inner eccentricity and mutual inclination to vary slightly, a family of plausible solutions can be obtained. Thus, if these preliminary results are confirmed, we may be seeing a case of Kozai cycles with tidal friction in action. Two eclipse events, unfortunately, are not enough to break the degeneracy between possible solutions and fully determine the extent of dynamics in the system.

From the measured period of the outer orbit, the timings of the two observed tertiary events and the photodynamical model, we can calculate the times of past eclipse events and predict the times of future eclipse events. The eclipse preceding the Q9 event should have occurred shortly before the beginning of the {\sl Kepler} mission (centered around BJD 2454907.25). The next event is expected to occur on May 14, 2015, motivating this brief communication. By observing this and other future events, we will be able to further constrain the model and determine the exact degree and nature of the tentative dynamical effects in the system. 

Our preliminary photodynamical models that fit each observed eclipse are depicted by a solid line in Fig.~2. By extrapolating both models to the time of the next eclipse event, we can obtain a reasonable estimate of the duration and shape of the upcoming event (Fig.~4). The exact times of ingress and egress, as well as the shape of the event, depend on the exact nature of these dynamical effects. It is thus important to observe the entire event so that a comprehensive model can be fully constrained.

\IBVSfig{8cm}{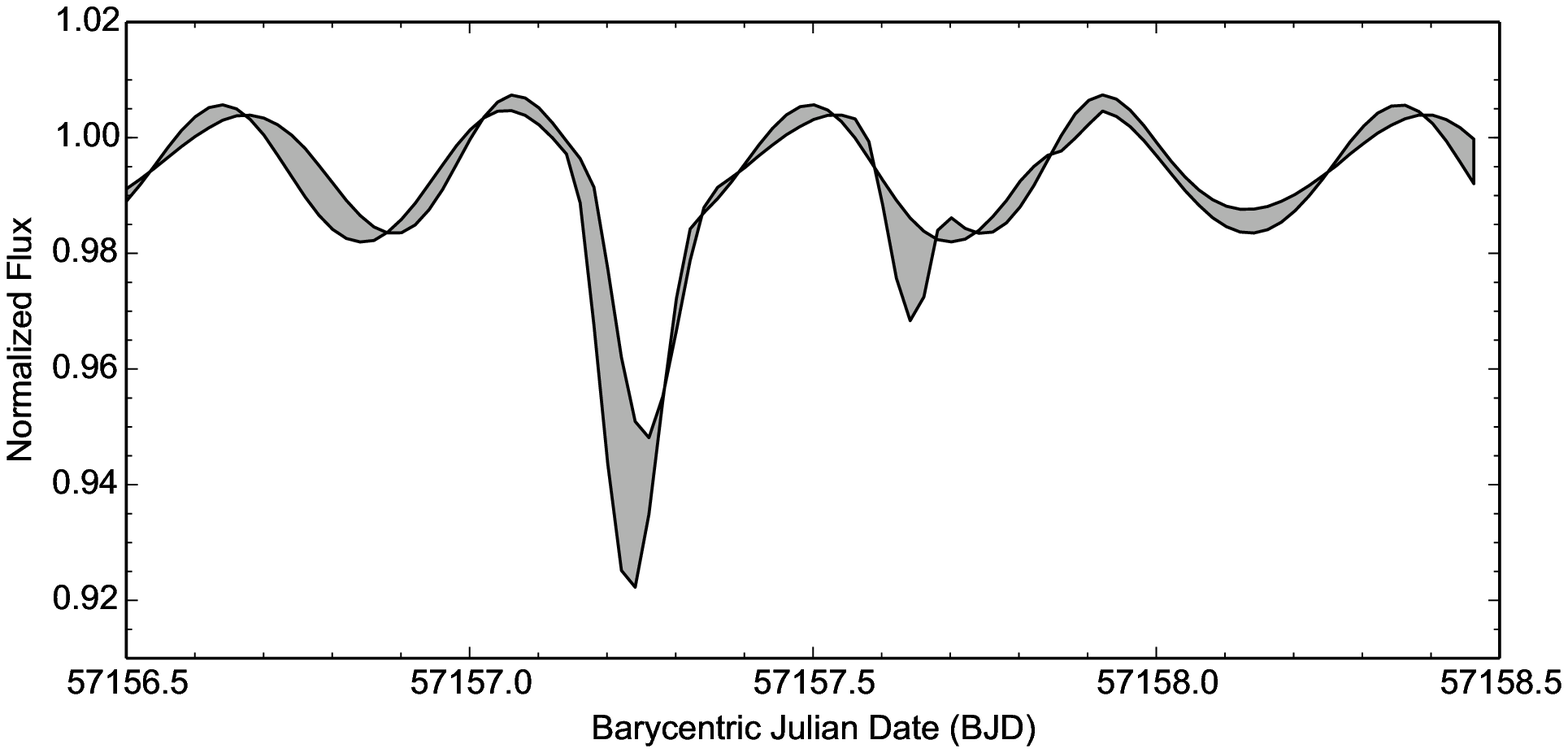}{
Light curve of the upcoming event. Solid lines depict predictions of the best-fit models to the Q9 eclipse event and the Q17 eclipse event. The shaded region represents the uncertainty of the prediction. This event is expected to occur on May 14: BJD 2457157.1--2457157.9.}

The next event is expected to occur on May 14, from Barycentric Julian Date 2457157.1 to 2457157.9. The amplitude of ellipsoidal variability is $\sim$2\% while the depth of the triple eclipse event is predicted to $\sim$5-10\%. Photometric follow up effort is being organized; Table 1 lists the time spans of several upcoming events and we solicit help from the community in observing them via this communication. For further coordination of this effort please contact the authors.

\begin{table}[!h]\label{table:timings}
\centering
\medskip
\begin{tabular}{rrrr}
\hline
Start Date (UTC) & BJD Start & BJD End \\
\hline
May 14, 2015 & 2457157.1 & 2457157.9 \\
June 2, 2017 & 2457907.2 & 2457908.3 \\
June 23, 2019 & 2458657.5 & 2458658.5 \\
July 12, 2021 & 2459407.5 & 2459408.5 \\
August 1, 2023 & 2460157.8 & 2460158.6 \\
\hline
\end{tabular}
\caption{Predicted earliest times of ingress and latest times of egress for the next several triple eclipse events by extending both the model to the Kepler Q9 and Q17 events.}
\end{table}

\newpage

\references

Borkovits, T., Rappaport, S., Hajdu, T., \& Sztakovics, J., 2015, {\it MNRAS}, {\bf 448}, 946. \BIBCODE{2015MNRAS.448..946B}

Borucki, W.J., et al., 2010, {\it Science}, {\bf 327}, 977. \BIBCODE{2010Sci...327..977B}

Conroy, K., et al., 2013, {\it EAS}, {\bf 64}, 295. \BIBCODE{2013EAS....64..295C}

Conroy, K., et al., 2014, {\it AJ}, {\bf 147}, 45. \BIBCODE{2014AJ....147...45C}

Fabrycky, D., \& Tremaine, S., {\it ApJ}, {\bf 669}, 1298. \BIBCODE{2007ApJ...669.1298F}

Kisseleva, L., Eggleton, P.P., \& Mikkola, S., 1998, {\it MNRAS}, {\bf 300}, 292. \BIBCODE{1998MNRAS.300..292K}

Pr\v sa, A., \& Zwitter, T., 2005, {\it ApJ}, {\bf 628}, 426. \BIBCODE{2005ApJ...628..426P}

Pr\v sa, A., et al., 2011, {\it AJ}, {\bf 141}, 83. \BIBCODE{2011AJ....141...83P}

\endreferences

\end{document}